\documentclass[11pt,twoside]{article}

\usepackage{times}

\usepackage{amsmath,amssymb,amsthm}
\usepackage{graphics,epsfig,calc}
\usepackage{amsmath}
\usepackage{latexsym,epsfig,bm,amssymb}
\usepackage{color}
\usepackage{amsthm,mathrsfs}

\usepackage{mathptmx}

\DeclareSymbolFont{AMSb}{U}{msb}{m}{n}
\DeclareSymbolFontAlphabet{\mathbb}{AMSb}

\usepackage{fancyhdr}
\newcommand{\beqn}{\begin{eqnarray}}
\newcommand{\eeqn}{\end{eqnarray}}
\newcommand{\be}{\begin{equation}}
\newcommand{\ee}{\end{equation}}
\newcommand{\ba}{\begin{array}}
\newcommand{\ea}{\end{array}}
\newcommand{\bo}{{\hfill\loota}}
\newcommand{\loota}{\hbox{\enspace{\vrule height 7pt depth 0pt width 7pt}}}

\newcommand{\bH}{{\bf H}}

\newcommand{\cD}{{\cal D}}
\newcommand{\cE}{{\cal E}}

\newcommand{\cL}{{\cal L}}

\newcommand{\cS}{{\cal S}}
\newcommand{\cT}{{\cal T}}

\newcommand{\al}{\alpha}

\newcommand{\ci}{\cite}
\newcommand{\de}{\delta}
\newcommand{\De}{\Delta}
\newcommand{\ds}{\displaystyle}
\newcommand{\fr}{\frac}

\newcommand{\la}{\label}
\newcommand{\lam}{\lambda}
\newcommand{\Lam}{\Lambda}
\newcommand{\na}{\nabla}

\newcommand{\ov}{\overline}

\newcommand{\pa}{\partial}
\newcommand{\re}{\ref}

\newcommand{\ti}{\tilde}
\newcommand{\ve}{\varepsilon}

\newcommand\R{{\mathbb R}}

\newcommand{\5}{{\hspace{0.5mm}}}

\newcommand{\3}{{\hspace{0.1mm}}}

\newcommand{\const}{\mathop{\rm const}\nolimits}

\newcommand{\tr}{\mathop{\rm tr\3}\nolimits}

\newcommand{\rRe}{{\rm Re\5}}
\newcommand{\rIm}{{\rm Im\5}}

\renewcommand{\Pr}{\hspace{-6mm}{\bf Proof~}}

\newtheorem{theorem}{Theorem}[section]

\newtheorem{definition}[theorem]{Definition}

\newtheorem{lemma}[theorem]{Lemma}
\newtheorem{example}[theorem]{Example}
\newtheorem{remark}[theorem]{Remark}
\newtheorem{remarks}[theorem]{Remarks}
\newtheorem{cor}[theorem]{Corollary}
\newtheorem{pro}[theorem]{Proposition}

\newcommand{\bd}{\begin{definition}}
 \newcommand{\ed}{\end{definition}}
\newcommand{\bt}{\begin{theorem}}
 \newcommand{\et}{\end{theorem}}
\newcommand{\bqt}{\begin{qtheorem}}
 \newcommand{\eqt}{\end{qtheorem}}

\newcommand{\bp}{\begin{pro}}
 \newcommand{\ep}{\end{pro}}

\newcommand{\bl}{\begin{lemma}}
 \newcommand{\el}{\end{lemma}}
\newcommand{\bc}{\begin{cor}}
 \newcommand{\ec}{\end{cor}}

\newcommand{\bex}{\begin{example}}
 \newcommand{\eex}{\end{example}}
\newcommand{\bexs}{\begin{examples}}
 \newcommand{\eexs}{\end{examples}}

\newcommand{\bexe}{\begin{exercice}}
 \newcommand{\eexe}{\end{exercice}}

\newcommand{\br}{\begin{remark} }
 \newcommand{\er}{\end{remark}}
\newcommand{\brs}{\begin{remarks}}
 \newcommand{\ers}{\end{remarks}}
\begin{document}

\begin{center}
{\huge\bf On the Hartree-Fock  dynamics \medskip\\
in wave-matrix picture}
\medskip\medskip\\
{\Large A.\,I.\,Komech}\footnote{
The research supported partly by Austrian Science Fund 
(FWF): P28152-N35, and the RFBR grant 13-01-00073.}
\medskip\medskip\\
{\it
Faculty of Mathematics of Vienna University
\medskip\\
Institute for Information Transmission Problems RAS}
\medskip\\
alexander.komech@univie.ac.at
%akomech@iitp.ru

\end{center}

\abstract{
We introduce the Hamiltonian dynamics 
with the Hartree-Fock energy
in  new  {\it wave-matrix} picture.
Roughly speaking, 
the wave matrix is defined as the square root of the density matrix.

The corresponding Hamiltonian equations are equivalent to an operator anticommutation equation.
This wave-matrix picture essentially agrees  with the  
density matrix formalism. Its main advantage is that it 
is Hamiltonian and 
allows an extension to infinite particle systems
like crystals in contrast with the  standard HF theory.

Our main result is the existence of the global "reduced" wave-matrix 
dynamics for finite-particle molecular  systems, and the energy and charge conservation laws.
For the proof we extend the techniques, 
based on Hardy's and Sobolev's inequalitites,
to the wave-matrix picture.

}
\bigskip


\setcounter{equation}{0}

\section{Introduction}
The first version of the  Hartree-Fock method 
was introduced by Hartree in 1927, and was refined by Fock and Slater about 1930 taking into account 
the antisymmetry of the fermionic wave functions. 
The method
is widely used in Quantum Chemistry 
for numerical determination of the ground state 
of finite particle molecular systems \ci{MUNRBG2006}. The main idea is the restriction of the test wave functions
in the Schr\"odinger minimization problem
to the set of the "Slater determinants".
The method is very efficient numerically and the 
results  are in a good agreement  with the corresponding experimental data.

The first rigorous results
on the existence of the ground state were established by 
Lieb and Simon \ci{LS1977} and by P.-L. Lions \ci{Lions1987}
for 
finite-particle molecular systems.
More general  {\it multiconfiguration} version of the Hartree-Fock theory has been  developed in 
 \ci{B1994,F2003,L2004}.
 \medskip

In 2001, the existence of the Hartree-Fock ground state has been established for crystals with 
space-periodic 
nuclei arrangements by Catto, Le Bris and  P.-L. Lions \ci{CBL2001}.
Next step should be 
an analysis of 
the dynamic properties of crystals near the 
ground state: its stability, dispersion, scattering theory, heat and electric conduction, etc.
However, the quantum dynamics of crystals is not rigorously established up to now.
For instance, the  rigorous
quantum theories of Ohm's Law and  Fourier's Law are missing 
\ci{BLR,Mad1996}
(see also the Preface  \ci{Peierls}).
\medskip

The rigorous time-dependent  Hartree-Fock theory
 has been developed first by Chadam and Glassey \ci{CG1975} for the reduced Hartree-Fock equations:
\be\la{Hi}
i\dot\psi_k(t)=H(t)\psi_k(t),~~~~~~~~~k=1,...,N;~~~~~~~\langle\psi_k(t), \psi_l(t) \rangle=\de_{kl}.
\ee
Here $\psi_k(t)\in L^2:=L^2(\R^3)$ for $t\in\R$, and 
$H(t):=-\De+eV_n(x)+eV_e(x,t)$
where $V_n(x)$ is the potential generated by the (standing) nuclei while  $V_e(x,t)$ is the potential generated by moving 
electrons:
\be\la{VcTi}
V_e(x,t)=\ds\int\fr{\rho(y,t)}{|x-y|}dy, ~~~~~~~~\rho(y,t):=e\sum_1^N|\psi_k(y,t)|^2\le 0,
\ee
where $e<0$ is the electron charge.
The well-posedness in the case of moving nuclei (Hellmann-Feynman nuclei dynamics) has been established 
by Canc\`es and Le-Bris \ci{CB1999}. 
The Hartree-Fock equations (\re{Hi}) are equivalent to the von Neumann equation
\be\la{etr2i}
i\dot K(t)=[H(t), K(t)]
\ee
for $K(t):=\sum_1^N|\psi_k(t)\rangle\langle\psi_k(t)|$. 
This equation can be considered for more general  {\it density matrices} $K(t)$ which are nonnegative selfadjoint trace class operators:
\be\la{dmtri}
 K^*(t)= K(t)\ge 0,~~~~ K(t)\le 1,~~~~\tr  K=N.
\ee
where the condition $K(t)\le 1$ corresponds to the Pauli exclusion principle, and $N$ is the "number of particles".
Now $H(t)$ is defined as above with $\rho(y,t):=eK(y,y,t)\le 0$. 

Dynamic of density matrices
(\re{etr2i})
was introduced initially
by von Neumann and Dirac about 1930 \ci{D1930,vN1996}, and  it
was used  in many cases. For example, in the superconductivity theory by 
Bogoliubov \ci{B1959} and
Valatin \ci{V1961}.
The  well-posedness for the von Neumann equation
was proved by 
Bove, Da Prato and Fano \ci{BPF1974,BPF1976} for 
a short-range pair-wise interaction potential $w(x-y)$ instead of the Coulomb potential $1/|x-y|$ in (\re{VcTi}).
The case of the Coulomb potential was solved 
by Chadam \ci{C1976}.
Butz and Spohn have applied the von Neumann equation with a source to phase transitions 
in the fermion/boson production \ci{BS2010}.
The multiconfiguration dynamics was constructed in \ci{BCMT2009}.
\medskip

However, the dynamical equation (\re{etr2i})
cannot be extended directly to infinite particle systems like crystals
since the corresponding Hamilton generator  is infinite: for example, the integral 
 (\re{VcTi}) diverges if $\rho(\cdot,t)$ is a space-periodic function.

 In \ci{CS-2012}, Cances and Stoltz have estabilshed the well-posedness  for 
local perturbations of the periodic ground state density matrix
in an infinite crystal in the  {\it random phase approximation}.
However, the  space-periodic nuclear potential 
in the equation \ci[(3)]{CS-2012}
is fixed that corresponds to 
the fixed nuclei positions. Thus the back reaction of the electrons onto 
the nuclei is neglected.
 
 The nonlinear Hartree-Fock dynamics
for 
compact perturbations of the ground state
without the  random phase approximation
was not studied previously,
see the discussion in 
\ci{BL2005} and in Introductions of \ci{CLL2012,CS-2012}.

In \ci{LS2014-1}, Lewin and Sabin have established the well-posedness for the 
von Neumann equation (\re{etr2i}) with density matrices of infinite trace 
for pair-wise interaction potentials $w\in L^1(\R^3)$. Moreover, the authors  
prove the asymptotic stability of the ground state in 2D case \ci{LS2014-2}.
The integral (\re{VcTi}) 
with  $w(x-y)$
instead of the Coulomb potential
obviously converges for  
$w\in L^1(\R^3)$ and
space-periodic functions $\rho(\cdot,t)$.
Let us stress however, that
the case of  the Coulomb potential  in \ci{LS2014-1} 
is not included.
\medskip

Thus a selfconsistent 
theory of the electron-lattice interaction is missing. 
A natural strategy 
to remedy the situation
would be the renormalization of the Hamilton functional by formal subtraction 
of infinite ground state energy. However, 
the Hamilton structure of the von Neumann equation
(\re{etr2i}) is not obvious (though the equations (\re{Hi})
are Hamiltonian, see (\re{DH}) below).
Hence, the theory requires a suitable Hamilton type modification.

Let us emphasize, that the Hartree-Fock 
dynamics is not canonically defined since the nonlinear 
manifold of the Slater determinants is not invariant with respect to the original 
Schr\"odinger dynamics. The relevance of the time-dependent Hartree-Fock equations
(\re{Hi})
is discussed in \ci[p.340]{BL2005}:
"The relation between the time-dependent Hartree-Fock equation and the original Schr\"odinger 
equation is mostly unclear (mathematically)". The results \ci{BGGM2003} justify
the relation "only for well-prepared initial states (Slater determinants, 
and slightly more general initial data), and only in the {\it weak coupling} picture",  see \ci[p.340]{BL2005}.
\medskip

\vspace{-4mm}

We introduce a modified  Hamilton 
dynamics with the Hamilton functional equal to
the Hartree-Fock energy in new `wave-matrix' picture.
The 
evolution for the 
corresponding density matrix agrees to some extent with the standard 
Hartree-Fock equations. 
Let us stress however, that this evolution
is not identical  with the 
Hartree-Fock equations, see Remark \re{rvN}.

The main advantage of  this wave-matrix
dynamics is that it
allows 
an extension to crystals by the renormalization
 of the Hamilton functional, as we will show elsewhere.
\medskip

In present paper we develop the wave-matrix theory for finite particle molecular systems.
  Our main result is the existence and uniqueness of global solutions 
 for the "reduced" wave-matrix dynamics.
 All estimates for operator-valued solutions are obtained in the corresponding
 Sobolev norms of their integral kernels.
 For the proof we extend the techniques 
 of the Hartree-Fock theory \ci{CB1999,CG1975,C1976,LS1977,Lions1987}, based on Hardy's and Sobolev's inequalitites,
 to the wave-matrices which are operator-valued functions.
 This extension is our main technical novelty (see Section 6).

 We establish the energy and charge conservation
 as well as all needed properties (\re{dmtri}) of the corresponding density matrix. 
 We check that for the molecular ground state
this wave-matrix picture  is equivalent to the standard
Hartree-Fock theory. 
Moreover, we show that the wave-matrix dynamics essentially
agrees   with the von Neumann equation (\re{etr2i}).
 \medskip

Our plan is the following.
In Sections 2 and 3 we recall the  Hartree-Fock theory
for the stationary and time-dependent cases.
In Section 4 we introduce the wave-matrix Hamilton equations and 
rewrite it as anticommutation equation.
In Section 5 we formulate our main result, and in 
Section 6 we establish needed technical estimates.
In Section 7 we reduce the dynamical equation  to the corresponding integral
Duhamel-type equation.
In Section 8
we construct local solutions, and 
Section 9 we prove the conservation laws. 
In Section 10 we obtain a priori bounds and construct global solutions.

In Section 11 we discuss the agreement of the wave-matrix Hamilton equation 
 with the Hartree-Fock 
density matrix formalism. 
Finally, in Appendix we calculate variational derivatives of the Hartree-Fock energy
in the wave-matrix picture.

\section{Hartree-Fock theory for ground state}
Let us recall  the  Hartree-Fock theory for
a molecule which consists of $M$ nuclei
with charges $|e|Z_j$. Let    $x^j\in\R^3$ denote the nuclei
locations, and $N=\sum_1^M Z_j$ the  number of the electrons. 
The Schr\"odinger dynamics for the molecule reads
\be\la{SdN}
i\dot\Psi(\ov x,t)=\bH\Psi(\ov x,t):=
-\sum_1^N\De_{x_k} \Psi(\ov x,t)+e[\sum_1^N V_n(x_k)+V_e(\ov x)]\Psi(\ov x,t),~~~~~\ov x\in\R^{3N}.
\ee
Here  $\ov x=(x_1,...,x_N)$, and 
\be\la{VV}
V_n(x):=\sum_1^M \fr {|e|Z_j}{|x-x_j|}, ~~~~~~V_e(\ov x):=\sum_{k<l} \fr {e}{|x_k-x_l|} 
\ee
are the potentials generated by the nuclei,
and the electrons  respectively. The wave function $\Psi(\ov x,t)$ is antisymmetric in $x_1,...,x_N$, and the ground state is the state $\Psi(x)$  with the minimal   Schr\"odinger energy
\be\la{gs}
E:=\min\{\ds\fr12\langle\Psi,\bH\Psi\rangle:~
\Vert\Psi\Vert_{L^2(R^{3N})}=1\}.
\ee
The Hartree-Fock method takes the minimum over the antisymmetric states of particular  form
 $\Psi(\ov x)=
\ds\fr1{\sqrt{N!}}\det \psi_k(x_l)$ (Slater determinant) 
with the constraints
\be\la{cons}
\langle\psi_k,\psi_l\rangle=\de_{kl}.
\ee
In this case the Schr\"odinger energy can be written as the Hartree-Fock functional  \ci{BJ,CG1975,LS1977,Lions1987} 
\beqn\la{ehf}
\fr12\langle\Psi,\bH\Psi\rangle=
\cE^{HF}(\Psi_N)&:=&\fr12 \sum_1^N \int |\na\psi_k(x)|^2dx+
\fr12 \int V_n(x)\rho(x)dx
\nonumber\\
\nonumber\\
&+&
\fr14\int 
\fr{\rho(x)\rho(y)}{|x-y|}dxdy
-\fr14 \int\int\fr{|\tau(x,y)|^2}{|x-y|}dxdy.
\eeqn
Here $\Psi_N:=(\psi_1,...,\psi_N)$, while
$\rho(x)$ is the electron charge density, and
\be\la{dm}
\rho(x)=eK(x,x),~~~\tau(x,y)=eK(x,y)~,~~~~~~~~~
K(x,y):=
\sum_1^N\psi_k(x)\ov{\psi_k(y)}.
\ee
 The 
{\it density matrix} $K$ 
is defined as the operator on the Hilbert space $X:=L^2(\R^3)$ with the integral kernel $K(x,y)$. It 
is the trace class nonnegative selfadjoint 
operator on $X$:
\be\la{dmtr}
 K^*= K\ge 0,~~~~K\le 1,~~~~\tr  K=N.
\ee
We keep throughout identical notations for operators and their integral kernels.
The energy $\cE^{HF}(\Psi_N)$ can be expressed in the
density matrix (\re{dm}) as
\be\la{etr}
\cE^{HF}(K)=-\fr 12 \tr \De  K+
\fr12 \int V_n(x)\rho(x)dx+
\fr14\int 
\fr{\rho(x)\rho(y)}{|x-y|}dxdy
-\fr14 \int\int\fr{|\tau(x,y)|^2}{|x-y|}dxdy.
\ee
Let us denote by $\cS(N)$ the set of all $\Psi_N=(\psi_1,...,\psi_N)\in \oplus_1^N X$ satisfying the constraints (\re{cons}). 
Then the Hatree-Fock approximation for the ground state energy (\re{gs}) reads
\be\la{gsHF}
E^{HF}:=\min\{\cE^{HF}(\Psi_N):~  \Psi_N\in \cS(N)\}\ge E.
\ee
Further, the density matrix $K=K(\Psi_N)=\sum|\psi_k\rangle\langle\psi_k|$ is invariant with respect to the 
unitary transformations 
\be\la{UN}
\Psi_N=(\psi_k: k=1,...,N)\mapsto \Phi_N=(\phi_k=\sum U_{kl}\psi_l: k=1,...,N),~~~~~(U_{kl})\in U(N).
\ee
Respectively, the Hartree-Fock energy  $\cE^{HF}$ also is $U(N)$-invariant functional. 
\medskip

The Hartree-Fock theory is widely used in quantum chemistry \cite{MUNRBG2006}.
Namely, the minimization of the energy (\re{ehf}) 
under the constraints (\re{cons})
provides a good approximation to
the molecular ground state energy (\re{gs}). The crucial advantage of this minimization problem 
is that it concerns $N$ functions of $3$ variables while the original Schr\"odinger problem (\re{gs}) concerns 
one function of $3N$ variables. However, the problem with $3N$ variables
is numerically unrealistic even for $N=10$ (as for the water molecule $H_2O$)
since the function of 30 variables with 20 points in each variable requires at least $20^{30}$ "cells" in memory, while
 $N$ functions of $3$ variable require $10\times 20^3$ cells.
\medskip

The Lagrange multipliers method 
leads to the variational equations 
\be\la{vare}
D_{\ov\psi_k}\cE^{HF}(\Psi_N)=\sum_{l=1}^N\lam_{kl}\psi_l,~~~~~~~~k=1,...,N.
\ee
Here $D_{\ov \psi_k(x)}:=D_{p_k(x)}+iD_{q_k(x)}$ where $p_k(x)=\rRe \psi_k(x)$ and 
 $q_k(x)=\rIm \psi_k(x)$.
Furthermore, the calculation gives that 
\be\la{DH}
D_{\ov\psi_k}\cE^{HF}(\Psi_N)=H\psi_k,
\ee
where $H=H(\Psi_N)$ is the symmetric operator in $X$ with the domain $\cD:=C_0^\infty(\R^3)$,
\be\la{Hs}
H(\Phi_N)=-\De+eV_n(x)+eV_e(x)+e\cT.
\ee
Here 
the potential generated by the electrons, $V_e(x)$,
and the operator $\cT$ are given by 
\be\la{VcTs}
V_e(x)=\ds\int\fr{\rho(y)}{|x-y|}dy,~~~~~~~~
\cT\psi(x)=-\int\fr{\tau(x,y)}{|x-y|}\psi(y)dy.
\ee
Now (\re{vare}) reads as
\ci[(12)]{Lions1987}
\be\la{vareH}
H(\Psi_N)\psi_k=\sum_{l=1}^N\lam_{kl}\psi_l,~~~~~~~~k=1,...,N.
\ee
Finally, the matrix $\Lam=(\lam_{kl})$ is Hermitian since
$H(\Psi_N)$ is the symmetric operator.
Hence, applying to  the both sides of (\re{vareH})
the matrix $U$, which diagonalize $(\lam_{kl})$, we obtain 
 \ci[(13)]{Lions1987}
\be\la{vare2}
H(\Phi_N)\phi_k=\ve_k\phi_k,~~~~~~~~k=1,...,N
\ee
since 
 $H(\Psi_N)$ is invariant with respect to the unitary transformations (\re{UN}).
\medskip

The  first results on existence of the ground state for 
finite-particle molecular systems
were established by 
Lieb and Simon \ci{LS1977} and  P.-L. Lions \ci{Lions1987}.
By Lieb's result \ci{Lieb1981} (see also \ci{Bach1992}),
the minimization of the energy $\cE^{HF}$ over
the Slater-type density matrices
is equivalent to its minimization over
general density matrices with integral kernel 
\be\la{dma}
K(x,y)=\sum\lam_n u_n(x) \ov{u_n(y)}~,~~~~\langle u_k,u_l\rangle=\de_{kl}
~,~~~~0\le \lam_n\le 1~,~~~\sum\lam_n=N.
\ee
In these notations the result \ci{Lieb1981} means that
\be\la{gsHFL}
E^{HF}=\min\{\cE^{HF}(K):~ K^*=K, ~0\le K\le 1,~\tr K=N\}.
\ee

\section{Hartree-Fock dynamics}

The structure of the stationary equations (\re{vare2}) suggests
the
dynamical Hartree-Fock equations considered in \ci{CB1999, CG1975}:
\be\la{H}
i\dot\psi_k(x,t)=H(t)\psi_k(\cdot,t),~~~~~~~~~k=1,...,N.
\ee
Here $H(t)=H(\Psi(t))$ is the operator of type (\re{Hs}):
\be\la{Ht}
H(t):=-\De+eV_n(x)+eV_e(x,t)+e\cT(t)
\ee
with the potential $V_e(x,t)$
and the operator $\cT(t)$  defined similarly to (\re{VcTs}):
\be\la{VcT}
V_e(x,t)=\ds\int\fr{\rho(y,t)}{|x-y|}dy,~~~~~~~~
\cT(t)\psi(x)=-\int\fr{\tau(x,y,t)}{|x-y|}\psi(y)dy,
\ee
where $\rho(y,t)$ and $\tau(x,y,t)$ correspond to the density matrix
$K(x,y,t):=
\sum_1^N\psi_k(x,t)\ov{\psi_k(y,t)}$.
The Hatree-Fock dynamics (\re{H}) can be  expressed via 
the density matrix as the von Neumann equation
\be\la{etr2}
i\dot K(t)=[H(t), K(t)],~~~
\ee
where $ K(t)$ is the operator with the integral kernel $K(x,y,t)$,
and $H(t)$ is the corresponding operator (\re{Ht}).

In \ci{CG1975}
the global solutions were constructed for the reduced Hartree-Fock equations 
(\re{H}) (i.e., with the operator (\re{Ht}) without the
 last term). In \ci{C1976} the result has been extended to the equation (\re{etr2})
 with general density matrices (\re{dma}).
In \ci{CB1999} the existence of global solutions 
has been proved for 
equation (\re{H})
coupled to the Newton equations for the nuclei.

\setcounter{equation}{0}
\section{Wave-matrix  picture}

The dynamic equations  (\re{H}) are Hamiltonian by (\re{DH}).
 On the other hand, the Hamilton structure of 
 general equation (\re{etr2}) is not obvious.
 
Let us recall, that the Hartree-Fock 
dynamics is not canonically defined since the nonlinear 
manifold of the Slater determinants is not invariant with respect to the original 
Schr\"odinger dynamics (\re{SdN}). 
We suggest a modified Hamiltonian dynamics for general density matrices
in the new picture
via ``wave-matrix'' operator $w$: 
\be\la{dmaw}
 K=ww^*,~~~~~\mbox{or equivalently,}~~~~~~K(x,y)=\int w(x,z)w^*(z,y)dz=\int w(x,z)\ov {w(y,z)}dz~,
\ee
where  $w(x,z)$ is the integral kernel of the
operator $w$. For example, we can take $w:=K^{1/2}$ for any density matrix (\re{dma}). The representation
is suggested by the 
 eigenfunction
expansions (\re{dm}) and (\re{dma}).
Obviously, $K$ is a trace class selfadjoint operator for any Hilbert-Schmidt operator $w$.
\medskip

Below we introduce the corresponding 
 dynamics for the wave matrices $w(x,y,t)$. We will show 
that this dynamics provides 
\medskip\\
i) The same ground state energy as constructed in  \ci{LS1977,Lieb1981,Lions1987}. 
\medskip\\
ii) All properties (\re{dmtr}) for the density matrix (\re{dmaw}) at any time $t\in\R$ once they hold
at $t=0$.
\medskip\\
iii)
The evolution for the corresponding density matrix $K(t)=w(t)w^*(t)$
which
agrees to some extent with equation (\re{etr2}), see Remark \re{rvN}.
\medskip\\
Let us note that we do not fix the number of particles which is equal to 
$\tr K$ and coincides with the Hilbert-Schmidt norm of $w$. 
We will show that this number is conserved along the  wave-matrix dynamics.
\medskip

First, we generalize the definition (\re{ehf}) of  the Hartree-Fock energy
for the wave-matrices:
\beqn\la{hfew}
\ti\cE^{HF}(w)&:=&\fr 14   \int \int [|\na_x w(x,y)|^2+|\na_y w(x,y)|^2]dxdy
+\fr e4   \int \int
[V_n(x)+V_n(y)]| w(x,y)|^2dxdy
\nonumber\\
\nonumber\\
&&+
\fr14\int \int 
\fr{\ti\rho(x)\ti\rho(y)}{|x-y|}dxdy
-\fr14\int\int\fr{|\ti\tau(x,y)|^2}{|x-y|}dxdy.
\eeqn
Here 
$\ti\rho(x)$ and $\ti\tau(x,y)$ are defined similarly to  (\re{dm}):
\be\la{dmm}
\ti\rho(x)=e\ti K(x,x),~~~\ti\tau(x,y)=e\ti K(x,y)~,~~~~~~~~~
\ti K(x,y):=\fr12
\int [w(x,z)\ov{w(y,z)}+\ov{w(z,x)}w(z,y)]dz.
\ee
In other words, $\ti K(x,y)$ is the integral kernel of the symmetric nonnegative operator 
\be\la{Kww}
\ti K=\fr12\{w,w^*\}
=\fr12[ww^*+w^*w]
%=\fr12[K+L],~~~~~~K=ww^*,~~~~L=w^*w.  
\ee
The energy (\re{hfew}) can be expressed in the density matrix $\ti K$ similarly to  (\re{etr}):
\be\la{etrw}
\ti\cE^{HF}(w)=-\fr 12 \tr \De  \ti K+
\fr12 \int V_n(x)\ti \rho(x)dx+
\fr14\int 
\fr{\ti\rho(x)\ti\rho(y)}{|x-y|}dxdy
-\fr14 \int\int\fr{|\ti\tau(x,y)|^2}{|x-y|}dxdy=\cE^{HF}(\ti K).
\ee
Note that the "Slater-type" density matrices
$K=\sum_1^N|\psi_k\rangle\langle\psi_k|$
admit representation (\re{dmaw}) with $w=K$
since $K^*=K$  and $K^2=K$ due to the constraints (\re{cons}). Hence, 
in this case 
\be\la{dmm2}
\ti K=w=K,~~~~~\ti\rho=\rho, ~~~~~~~~\ti\tau=\tau,~~~~~~~\ti\cE^{HF}(w)=\cE^{HF}(K).
\ee
In this ``wave-matrix'' representation we accept $\ti\cE^{HF}(w)$ as the Hamilton functional.
 Respectively, 
we define the Hamilton wave-matrix dynamics formally by
\be\la{HFNdw}
i\dot w(x,y,t)=2D_{\ov w(x,y)}\ti \cE^{HF}(w(\cdot,t))=-[\De_x+\De_y]w(x,y,t)+ ...
\ee
Here
$D_{\ov w(x,y)}:=D_{w_1(x,y)}+iD_{w_2(x,y)}$ where $w_1(x,y)=\rRe w(x,y)$ and 
 $w_2(x,y)=\rIm w(x,y)$.
 We change the "standard" Hamilton structure
 introducing the prefactor $2$,
 to reconcile the dynamics with the von Neumann equation (\re{etr2})
 as we will show later.
\medskip

Let us denote by   $\ti H(t)$ the operator (\re{Ht}) with 
 the potential $V_e(x,t)$ and the operator $\cT(t)$  
 changed to the corresponding $\ti V_e(x,t)$ and $\ti \cT(t)$  
 which are
 defined similarly to (\re{VcT}):
 \be\la{tHt}
\ti H(t):=-\De+eV_n(x)+e\ti V_e(x,t)+e\ti\cT(t),
\ee
\be\la{tiVt}
\ti V_e(x,t)=\ds\int \fr{\ti\rho(y,t)}{|x-y|}dy,~~~~~~~~~
\ti \cT(t)\psi(x)=-
\int\fr{\ti\tau(x,y,t)}{|x-y|}\psi(y)dy.
\ee
Formally calculating the variational derivative in  (\re{HFNdw}),
we obtain  
\be\la{HFw}
i\dot w(t)=\{\ti H(t),w(t)\}:=\ti H(t)w(t)+w(t)\ti H(t).
\ee
We justify this calculation in Lemma \re{lGD} for the {\it reduced} equation (\re{HFNdw}).
\medskip

\setcounter{equation}{0}

\section{Wave-matrix dynamics}
We will prove the existence and uniqueness of global solutions to the reduced equation (\re{HFw})
when the operator $\ti H(t)$ is defined by (\re{tHt}) without $\ti \cT(t)$. In other words, from now on,
\be\la{tHt2}
\ti H(t):=H_0+\ti V(t),~~~~~~~~~~H_0:=-\De+eV_n(x)~~~~~~\ti V(t):=e\ti V_e(x,t).
\ee
Respectively, the Hamiltonian $\cE^{HF}(w)$ now is changed to the reduced Hartree-Fock energy
\be\la{etrwr}
\ti\cE^{RHF}(w)=\fr 12 \tr [ H_0 \ti K(t)]+
\fr14\int 
\fr{\ti\rho(x)\ti\rho(y)}{|x-y|}dxdy,
\ee
and the corresponding dynamic equation 
(\ref{HFNdw})
formally reads
\be\la{HRFNw}
i\dot w(t)=2D_{\ov w}\ti \cE^{RHF}(w(\cdot,t)),~~~~~~~~~~~~t\in\R.
\ee
To formulate our main results we need the following definition.
Let us denote by $\cL^2$ the Hilbert space of the Hilbert-Schmidt operators in $L^2$.

\bd\la{dHs}
$\bH^s$ with $s=0,1,...$ denotes the space of operators $w\in\cL^2$ endowed 
with the finite
norm
\be\la{bHs}
\Vert w\Vert_{\bH^s}^2=\sum_{|\al|\le s}\int|\pa^\al_{(x,y)} w(x,y)|^2dxdy,
\ee
where $w(x,y)$ denotes the integral kernels of $w$. 
\ed
Equivalently,
$\pa^\al_x w\in \cL^2$ and $w \pa^\al_x\in \cL^2$ for $|\al|\le s$.
In particular, $\bH^0=\cL^2$. 
\medskip

We will construct strong solutions $w(\cdot)\in X:=C^1_s(\R,\bH^0)\cap C(\R,\bH^2)$,
where $C^1_s$ denotes the strongly differentiable operator functions, while 
$C(\R,\bH^2)$ denotes the space of continuous operator functions in the  norm $\bH^2$.
In this case the equation (\re{HRFNw}) can be written as
\be\la{RHFw}
i\dot w(t)=\{\ti H(t),w(t)\},~~~~~~~~t\in\R
\ee
by the following lemma.
\bl\la{lGD}
The Hamilton functional $\ti\cE^{RHF}$ is G$\hat a$teaux differentiable on the space $\bH^2$, and  
\be\la{GD}
2D_{\ov w}\ti \cE^{RHF}(w)=\{\ti H,w\}\in\bH^0,~~~~~~~~~w\in\bH^2,
\ee
where $\ti H=H_0+e\ti V_e(x)$ and $\ti V_e(x):=\ds\int\fr{\ti\rho(y)}{|x-y|}dy$.
\el
We  prove this lemma in Appendix.
Our main result is the following theorem.

\bt\la{t1}
For any initial state $w(0)\in\bH^2$ there exists the unique strong solution 
$w(\cdot)\in X$ to (\re{RHFw}).
\et

For the proof we follow the standard 
scheme:
first we
prove some technical estimates and construct the local solutions; 
afterwords, we prove a priori
estimates which give the global strong solutions.
\medskip

In conclusion, let us differentiate the density matrix $K(t):=w(t)w^*(t)$ for a solution $w(\cdot)\in X$ to
(\re{RHFw}).
Taking the 
adjoint to (\re{RHFw}), we get
$-i\dot w^*(t)=w^*(t)\ti H(t)+\ti H(t)w^*(t)$, and hence,
\be\la{HFwK}
i\dot  K(t)=i[\dot w(t) w^*(t)+w(t)\dot w^*(t)]=[\ti Hw+w \ti H]w^*-w[w^*\ti H+\ti Hw^*]=[\ti H(t), K(t)].~~~~~~
\ee
\br\la{rvN}
Equation 
(\re{RHFw}) for the wave matrix $w(t)$ agrees with the von Neumann
equation (\re{etr2}) for  $K(t):=w(t)w^*(t)$ 
at such times $t$  that $w(t)$ is the Slater-type 
 density matrix. Namely,  $\ti H(t)=H(t)$ for these times by  (\re{dmm2}), and hence 
 (\re{HFwK}) coincides with (\re{etr2}) for these times.
 Let us stress however, that the evolution of $K(t)=w(t)w^*(t)$
is not identical to (\re{etr2}).
 \er

\setcounter{equation}{0}

\section{Basic estimates}
We 
extend basic estimates \ci{CB1999} 
to the wave-matrix formalism.
First let us obtain estimates for the potential $\ti V_e$ defined in (\re{tiVt}):
\be\la{Vce}
\ti V_e(x)=\ds\int\fr{\ti\rho(y)}{|x-y|}dy=\fr12 \int \fr{\ds\int[w(y,z)\ov w(y,z)+\ov w(z,y)w(z,y)]dz}{|x-y|}dy.
\ee

\bl\la{l1} Let $w\in \bH^1$. Then 
\be\la{Ve}
\sup_{x\in\R^3}|\ti V_e(x)|\le C\Vert w\Vert_{\bH^0}\Vert w\Vert_{\bH^1}.
\ee

\el
\Pr  Let us denote the integrand
\be\la{Vce1}
\ti V_e(x,z)=\fr12 \int \fr{w(y,z)\ov w(y,z)+\ov w(z,y)w(z,y)}{|x-y|}dy.
\ee
Applying 
the Cauchy-Schwarz and Hardy inequality \ci[p.446]{CH1989}, we obtain  
\be\la{Ve2}
|\ti V_e(x,z)|\le 
C(\Vert w(\cdot,z)\Vert\cdot\Vert \na_1 w(\cdot,z)\Vert+\Vert w(z,\cdot)\Vert\cdot\Vert \na_2 w(z,\cdot)\Vert),
\ee
where $\Vert \cdot\Vert$ denotes the norm in $L^2$, and $\na_1$,  $\na_2$ are obvious notations.
Now the integration over $z\in\R^3$ gives 
(\re{Ve}) by the Cauchy-Schwarz inequality.\bo
\medskip\\
This lemma implies that the anticommutator
\be\la{Fw}
F(w)=\{\ti V_e,w\}
\ee
is the Hilbert-Schmidt operator on $L^2$ 
for $w\in \bH^1$.
The anticommutator $\{\De,w\}$ is 
 the Hilbert-Schmidt operator
for $w\in \bH^2$. Finally, 
$\{V_n,w\}$ is the Hilbert-Schmidt operator  for $w\in \bH^1$
by the Hardy inequality.
As the result, the right hand side of 
(\re{RHFw}) is well defined Hilbert-Schmidt operator for $w(\cdot)\in X$.
\medskip\\
Further we need the local Lipschitz continuity  for the anticommutator (\re{Fw}).
Next two lemmas  extend  
Lemma 5 of \ci{CB1999}
to the wave-matrix formalism. The first lemma concerns the Lipschitz continuity in $\bH^0$, and the second one 
- in $\bH^2$.

 \bl\la{lF}
 {\rm (cf. Lemma 5 (a) of \ci{CB1999})}
For $w,w'\in H^1$
 \be\la{F}
 \Vert F(w)-F(w')\Vert_{\bH^0}\le  C(\Vert w\Vert_{\bH^1}^2+\Vert w'\Vert_{\bH^1}^2)\Vert w-w'\Vert_{\bH^0}.
 \ee
 \el
 \Pr
 It suffices to prove (\re{F}) for one term 
 \be\la{Fw2}
F_1(w)=w\int \fr{\ds\int w(y,z)\ov w(y,z)dz}{|x-y|}dy
\ee
since the proof for the other term is similar.
%Applying the arguments from \ci[Lemma 5]{CB1999}, we obtain
Obviously,
\be\la{Fw3}
F_1(w)-F_1(w')=(w-w')\int \fr{\ds\int w(y,z)\ov w(y,z)dz}{|x-y|}dy+
w'\int \fr{\ds\int [w(y,z)\ov w(y,z)-w'(y,z)\ov w'(y,z)]dz}{|x-y|}dy. 
\ee
The first term on the right hand side admits the bound (\re{F}) by previous lemma. For the 
second term  we estimate the "integrand" as in (\re{Ve2}):
\beqn\la{Fw4}
I(x,z)&=&\int \fr{w(y,z)\ov w(y,z)-w'(y,z)\ov w'(y,z)}{|x-y|}dy
\nonumber\\
\nonumber\\
&=&
\int \fr{w(y,z)\ov w(y,z)-w'(y,z)\ov w(y,z)+w'(y,z)\ov w(y,z)-w'(y,z)\ov w'(y,z)}{|x-y|}dy
\nonumber\\
\nonumber\\
&\le&
C(\Vert \na_1 w(\cdot,z)\Vert\cdot \Vert w(\cdot,z) -w'(\cdot,z)\Vert+
\Vert \na_1 w'(\cdot,z)\Vert\cdot \Vert w(\cdot,z) -w'(\cdot,z)\Vert).
\eeqn
Now the Cauchy-Schwarz inequality implies
\be\la{Fw5}
\sup_{x\in\R^3}\int|I(x,z)|dz\le C(\Vert w\Vert_{\bH^1}+\Vert w'\Vert_{\bH^1})\Vert w-w'\Vert_{\bH^0}.
\ee
Hence, the second term on the right hand side
of (\re{Fw3}) also
admits the bound (\re{F}).\bo
\medskip\\
Next lemma extends these estimates to $\bH^2$ norms.
\bl\la{lF2}
{\rm (cf. Lemma 5 (b) of \ci{CB1999})}
For $w,w'\in H^2$
\beqn
\la{Fw6}
\Vert F(w)\Vert_{\bH^2}  &\le& C_F  \Vert w\Vert_{\bH^1}^2   \Vert w\Vert_{\bH^2},
\\
\nonumber\\
\Vert F(w)-F(w')\Vert_{\bH^2}&\le & C_F(\Vert w\Vert_{\bH^2}^2+\Vert w'\Vert_{\bH^2}^2)\Vert w-w'\Vert_{\bH^2}.
\la{Fw7}
\eeqn

\el
\Pr  i) To prove (\re{Fw6}) we should bound the norms $\Vert F_1(w)\Vert_{\bH^0}$, $\Vert \De F_1(w)\Vert_{\bH^0}$, and $\Vert F_1(w)\De\Vert_{\bH^0}$. The first and second norms are bounded similarly to Lemma \re{lF}.
It remains to bound the third norm.
Let us consider the integrand of  (\re{Fw2}):
\be\la{Fw2z}
F_z=w\int \fr{ w(y,z)\ov w(y,z)}{|x-y|}dy. 
\ee
This is the integral operator with the kernel
\be\la{Fw2zk}
F_z(x,y')=w(x,y')\int \fr{ w(y,z)\ov w(y,z)}{|y'-y|}dy
%=w(x,y')\int \fr{ w(y'-y,z)\ov w(y'-y,z)}{|y|}dy. 
\ee
Further, $F_z\De$ is the integral operator with the kernel $\De_{y'}F_z(x,y')=\De_2F_z(x,y')$.
Now we differentiate (cf.  Lemma 5 of \ci{CB1999}):
\beqn\la{Fw2z2}
\!\!\!\!\!\!\!\!\!\!\De_2 F_z(x,y')\!\!\!\!&\!\!\!\!=\!\!&\!\!\!\!4\pi w(x,y')\cdot w(y',z)\ov w(y',z)   
\nonumber\\
\nonumber\\
\!\!\!\!\!\!\!\!\!\!\!\!\!\!&\!\!\!\!\!\!+\!\!&\!\!\!\!2\na_2 w\int \fr{ \na_1 w(y,z)\ov w(y,z)}{|x-y|}dy+2\na_2 w\int \fr{ w(y,z)\na_1\ov w(y,z)}{|x-y|}dy+\De_2 w\int \fr{ w(y,z)\ov w(y,z)}{|x-y|}dy. 
\eeqn
Here the first term on the right hand side is the operator with the integral kernel
\be\la{Fwabc}
K_z(x,y')=
4\pi w(x,y') w(y',z)\ov w(y',z).
\ee
Let us bound its Hilbert-Schmidt norm  
extending estimate (10) of \ci{CB1999} to the wave-matrix formalism:
applying the H\"older inequality and the Sobolev embedding theorem, we obtain that
\beqn\la{Fwabc2}
\int |K_z(x,y')|^2dy'&\le&
C\Vert w(x,\cdot)\Vert_{L^6}^2 \Vert w(\cdot,z)\Vert_{L^6(\R^3)}^2\Vert\ov w(\cdot,z)\Vert_{L^6(\R^3)}^2
\nonumber\\
\nonumber\\
&\le&
C\Vert w(x,\cdot)\Vert_{H^1(\R^3)}^2 \Vert w(\cdot,z)\Vert_{H^1(\R^3)}^2\Vert\ov w(\cdot,z)\Vert_{H^1(\R^3)}^2.
\eeqn
Integrating over $x\in\R^3$ we obtain 
\be\la{Fwabc4}
\Vert K_z\Vert_{\bH^0}\le
C\Vert w\Vert_{\bH^1(\R^3)} \Vert w(\cdot,z)\Vert_{H^1(\R^3)}\Vert\ov w(\cdot,z)\Vert_{H^1(\R^3)}.
\ee
Finally, integrating over  $z\in\R^3$ we obtain  by the Cauchy-Schwarz the bound (\re{Fw6}) for the contribution of the first term on the right hand side (\re{Fw2z2}). The bounds for the other three terms can be obtained by the same Cauchy-Schwarz
trick  using the bounds of type  (\re{Ve2}) for the corresponding integrals.
\medskip\\
ii) It suffices to prove (\re{Fw7}) for $F_1$. Obviously, 
\be\la{Fw71}
\Vert F_1(w)-F_1(w')\Vert_{\bH^2}\sim \Vert F_1(w)-F_1(w')\Vert_{\bH^0}+\Vert \De[F_1(w)-F_1(w')]\Vert_{\bH^0}+\Vert [F_1(w)-F_1(w')]\De\Vert_{\bH^0}.
\ee
 The first term on the right hand side is estimated by (\re{F}). For the second term the estimate follows from 
 (\re{Fw3}) by the same arguments (\re{Fw4})--(\re{Fw5}). 
 Finally, the estimate for the last term follows by the combination of 
the arguments  (\re{Fw4})--(\re{Fw5}) with the proof of  (\re{Fw6}) above.\bo

  \section{Integral Duhamel equation}
  
Let us 
 reduce  (\re{RHFw}) with $w(\cdot)\in X$ to an equivalent integral equation.
Using notations (\re{tHt2}), we rewrite (\re{RHFw}) as
\be\la{HFV}
i\dot w(t)=\{H_0,w(t)\}+\{ \ti V(t),w(t)\}.
\ee
We reduce this equation
to the case of  bounded generator withdrawing its unbounded part. Namely,
let us write the solution in the "interaction picture"
\be\la{wUC}
w(t)=U_0(t)C(t)U_0(t),~~~~~~~~~~t\in\R,
\ee
where $U_0(t):=\exp(-iH_0t)$ is the dynamical group of the "free" Schr\"odinger equation.
Obviously, $C(\cdot)\in X$ since $w(\cdot)\in X$. Hence, 
the differentiation gives
\be\la{wUC2}
\dot w(t)=\{H_0,w(t)\}+U_0(t)\dot C(t)U_0(t).
\ee
Substituting into (\re{RHFw}), we obtain the equivalent reduced equation 
\be\la{HFr}
 iU_0(t)\dot C(t)U_0(t)=\{\ti V(t),w(t)\}.
\ee
The integration gives
\be\la{HFr2}
 C(t)=C(0)-i\int_0^tU_0(-s) \{\ti V(s),w(s)\} U_0(-s)ds.
\ee
Coming back to $w(t)$, we get the  integral "Duhamel" equation
\be\la{HFr3}
 w(t)=U_0(t)w(0)U_0(t)-i\int_0^t U_0(t-s)\{\ti V(s),w(s)\}U_0(t-s)ds,~~~~~~~~~t\in\R.
\ee
\bl\la{lequiv}
For $w(\cdot)\in X$ the differential equation  (\re{RHFw}) is equivalent to its integral version
(\re{HFr3}).
\el
 \Pr To deduce (\re{HFr}) from (\re{HFr2}) for
 $w(\cdot)\in X$ it suffices to note that 
  the integrand  belongs to $C_s(\R,\bH^2)$ 
  (strongly continuous operator functions)
  since
  \be\la{tHwV2}
  \{\ti V(t),w(t)\}\in C(\R,\bH^2)
  \ee
 by (\re{Fw7}).\bo

  \section{Local solutions}
  
  Let us  prove that the local solution  exists  by the 
  Picard fix point theorem   due to the Lipschitz continuity. 
  Let us denote 
  $X_\ve:= C^1_s(-\ve,\ve;\bH^0)\cap C(-\ve,\ve;\bH^2)$ for $\ve>0$.
  
  \bl\la{lls}
  For any $w(0)\in\bH^2$ there exists a unique strong solution $w(\cdot)\in X_\ve$
  to the equation  
  (\re{RHFw}) for $|t|<\ve$ with  $\ve=\ve(C_F,\Vert w(0)\Vert_{\bH^2})>0$.
  \el
  \Pr 
  Operators $U_0(t)$ are uniformly bounded in $H^2(\R^3)$. Hence,
  due to (\re{Fw6}) and (\re{Fw7}) 
  the unique solution $w(\cdot)\in C(-\ve,\ve;\bH^2)$
  to the integral equation 
  (\re{HFr3}) exists by the Picard fix point theorem 
  for $|t|<\ve$ with  $\ve=\ve(C_F,\Vert w(0)\Vert_{\bH^2})>0$ (see \ci{S1963}).
  It remains to  prove that
  \be\la{wC1}
  w(\cdot)\in C^1_s(-\ve,\ve;\bH^0).
  \ee
  Indeed,  let us consider both terms on the right hand side of   (\re{HFr3}). The first term belongs to 
   $C^1_s(-\ve,\ve;\bH^0)$
  since 
  \be\la{wC12}
  \Vert \dot U_0(t)w(0)U_0(t)\Vert_{\bH^0}+\Vert U_0(t)w(0)\dot U_0(t)\Vert_{\bH^0}\sim 
  \Vert U_0(t)\ti H_0w(0)U_0(t)\Vert_{\bH^0}+\Vert U_0(t)w(0)\ti H_0U_0(t)\Vert_{\bH^0},
  \ee
  where $\ti H_0w(0)\in\bH^0$ and  $w(0)\ti H_0\in\bH^0$.
  Finally, 
   the integrand (\re{HFr3}) 
   belongs to $C^1_s(-\ve,\ve;\bH^2)$ since
  \be\la{tHwV}
  \{\ti V(t),w(t)\}\in C(-\ve,\ve;\bH^2)
  \ee
 by (\re{Fw7}).
 Hence, $w(\cdot)\in X_\ve$, and 
 (\re{HFr3}) 
 implies (\re{RHFw}) for $|t|<\ve$.\bo

 \section{Conservation laws}
 To deduce Theorem \re{t1} from Lemma \re{lls}, we need a priori estimates which follow 
 from energy and norm conservation.

\subsection{Energy conservation}
Let us prove the energy conservation
\be\la{enec}
\ti\cE^{HF}(w(t))=\const,~~~~~~~~~~~~~~~t\in\R.
\ee
 Formally, the conservation follows 
 by direct differentiation
 from the Hamilton structure of the equation 
(\re{HFNdw}). However, the formal differentiation cannot be justified with the
application of the standard chain rule due to a
mismatch in the estimates for the remainder. This is why we justify the differentiation directly using the polynomial
structure of the Hamilton functional.

\bl\la{lenec}
Let  $w(\cdot)\in X$ be a strong solution to (\re{RHFw}). Then  the energy conservation (\re{enec}) holds.
\el
\Pr Let us write the reduced energy (\re{etrwr}) for the solution $w(t)$ as
\be\la{etrwr2}
\ti\cE^{RHF}(w(t))=\fr 14 \tr[H_0  w(t)w^*(t)+ H_0 w^*(t)w(t)]+
\fr14\int \fr{\ti\rho(x,t)\ti\rho(y,t)}{|x-y|}dxdy.
\ee
Here the operators $H_0$, $w(t)$ and $w^*(t)$ can be cyclically permuted.
Hence, the derivative  can be written formally as
\beqn\la{ecd}
\dot{\ti \cE}^{RHF}(w(t))&=&
\fr 14 \tr[w^*(t)H_0  \dot w(t)+ H_0  w(t)\dot w^*(t)+w(t)H_0 \dot w^*(t)+H_0 w^*(t)\dot w(t)]
+\fr 12\langle \ti V_e(x,t),\dot{\ti\rho}(x,t)\rangle
\nonumber\\
\nonumber\\
&=&
\fr 14 \tr[\{H_0,w(t)\}  \dot w^*(t)+ \{H_0,  w^*(t)\}\dot w(t)]
+\fr 12\langle \ti V_e(x,t),\dot{\ti\rho}(x,t)\rangle.
\eeqn
To justify this differentiation, we first show that all these terms exist.
The terms with $H_0$ exist because $\dot w(t)\in\bH^0$, and also $\{H_0,w(t)\}\in\bH^0$
since  $w(t)\in\bH^2$. The last term
can be written similarly,
\beqn\la{ecd2}
\fr 12\langle \ti V_e(x,t),\dot{\ti\rho}(x,t)\rangle
&=&\fr14\tr[\ti V(t)(\dot w(t)w^*(t)+  w(t)\dot w^*(t)+\dot w^*(t)w(t)+w^*(t)\dot w(t))]
\nonumber\\
\nonumber\\
&=&\fr14\tr[\{\ti V(t),w(t)\}\dot w^*(t)+  \{\ti V(t),w^*(t)\}\dot w(t)].
\eeqn
This expression is finite since $\ti V(t)$ is the operator of multiplication by $e\ti V_e(\cdot,t)$ which is 
the bounded function by (\re{Ve}).

Now we can justify the differentiations (\re{ecd}). Since the energy is the fourth order polynomial 
in $w(t)$ and  $w^*(t)$,
the  increment 
$\De\ti \cE^{RHF}(t):=\ti \cE^{RHF}(w(t+\De t))-\ti \cE^{RHF}(w(t))$ can be written as the corresponding 
polynomial in  $w(t), w^*(t)$, and $\De w(t):=w(t+\De t)-w(t)$. The main part, linear in   $\De w(t)$,
looks like   (\re{ecd})-(\re{ecd2})
with $\dot w(t)$ substituted by $\De w(t)$ and  $\dot w^*(t)$ substituted by $\De w^*(t)$.
It remains to divide $\De\ti \cE^{RHF}(t)$
by $\De t$  and send  $\De t\to 0$.  
Then the contribution of the main part gives (\re{ecd})-(\re{ecd2}) by previous arguments. 
The contributions of the 
higher order terms
converge to zero by similar arguments.

Finally, let us prove that the derivative  (\re{ecd}) vanishes using the dynamic equation (\re{RHFw}).
First let us rewrite  
(\re{ecd})-(\re{ecd2}) as 
\be\la{ecd3}
\!\dot{\ti \cE}^{RHF}(w(t))\!=\!
\fr 14 \tr[\{H_0+\ti V(t), w^*(t)\}  \dot w(t)
+ \{H_0 +\ti V(t),w(t)\}\dot w^*(t)].
\ee
Substituting here $\dot w(t)=-i\{H_0+\ti V(t),w(t)\}$ and  $\dot w^*(t)=i\{H_0+\ti V(t),w^*(t)\}$,
we obtain zero since $w(t)\in\bH^2$, and hence both anticommutators $\{H_0+\ti V(t),w(t)\}$ and $\{H_0+\ti V(t),w^*(t)\}$ are the Hilbert-Schmidt operators.\bo

\subsection{Charge conservation}
Now we can prove  the charge conservation:
\be\la{cc}
Q(t):=\int\ti\rho(x,t)dx=\const,~~~~~~~~~~~~~~~~~~~~~~~~~t\in\R.
\ee
\bl\la{lcc} 
Let $w(\cdot)\in X$ be a strong solution to (\re{RHFw}). 
 Then the charge conservation (\re{cc}) holds.
 \el
\Pr  First, we note that $Q(t)=e\tr \ti K(t)=
e\tr w(t)w^*(t)=e\tr C(t)C^*(t)$ by
(\re{wUC})
since the operators $U_0(t)$ are unitary. So it remains to prove the 
conservation of $\tr C(t)C^*(t)$ which follows by the differentiation. Namely,
(\re{HFr}) implies
\be\la{Hfr2}
i\dot C(t)=V_L(t)C(t)+C(t) V_R(t),~~~~~~~~ V_L(t)=U_0^*(t)\ti V(t)U_0(t),~~~~~~~V_R(t)=U_0(t)\ti V(t)U_0^*(t).
\ee
Here the selfadjoint operators $V_L(t),V_R(t)\in C_s(\R,\cL)$ by the bounds of type (\re{Ve}) for differences
$\ti V_e(x, t+\De t)-\ti V_e(x,t)$, where $\cL=\cL(L^2,L^2)$ is the space of bounded operators in $L^2$,
and $C_s(\R,\cL)$ denotes the space of strongly continuous operator functions. 
Taking the adjoint to both sides, we obtain 
$-i\dot C^*(t)=C^*(t)V_L(t)+V_R(t)C^*(t)$, and hence 
\be\la{CC}
i\fr d{dt}[C(t)C^*(t)]=[V_L(t), C(t)C^*(t)].
\ee
Therefore, $\tr C(t)C^*(t)=\const$ since the trace of the commutator vanishes.\bo
\medskip

\subsection{Norm conservation}
Let us note that (\re{cc}) means that   $\Vert w(t)\Vert_{\bH^0}=\const$.
Further we will prove also the conservation of the operator norm in $\cL$:
\be\la{nc}
\Vert w(t)\Vert=\const,~~~~~~~~~~~~~~~~~~~~~t\in\R.
\ee
\bp\la{pnc} 
Let $w(\cdot)\in X$ be a strong solution to (\re{RHFw}). 
 Then the norm conservation (\re{nc}) holds.
\ep
\Pr
For the proof we need the following lemma.
\bl\la{lUU}
Let $w(\cdot)\in X$ be a strong solution to (\re{RHFw}), and  $V_L(t)$, $V_L(t\in C(\R,\cL)$ are the corresponding 
selfadjoint 
operators (\re{Hfr2}). 
Then
\medskip\\
i)
There exist unique  unitary propagators $U_L(t,s)$ and $U_R(t,s)$ 
which are solutions to
\beqn\la{UU}
i\dot U_L(t,s)&=&V_L(t)U_L(t,s),~~~~~~~~~t,s\in\R;~~~~~U_L(s,s)=I\\
i\dot U_R(t,s)&=&U_R(t,s)V_R(t),~~~~~~~~~t,s \in\R;~~~~~U_R(s,s)=I,
\eeqn
where the derivatives are understood in the strong sense.
\medskip\\
ii)
 The "group identities" hold
\be\la{Utsr}
U_L(t,s)U_L(s,r)=U_L(t,r),~~~~U_R(t,s)U_R(s,r)=U_R(t,r),~~~~~~~~~~~t,s,r\in\R.
\ee

\el
\Pr
The solutions exist 
and are unique
since $V_L(t),V_R(t)\in C(\R,\cL)$.
The identity 
(\re{Utsr}) holds by the uniqueness of the solutions.

The 
propagators are unitary operators since the generators $V_L(t)$, $V_R(t)$ are selfadjoint.
For example, the adjoint equation to (\re{UU}) reads $\dot U_L^*(t,s)=iU_L^*(t,s) V_L(t)$, and hence
\beqn\la{oeq2}
\fr d{dt} [U_L^*(t,s)U_L(t,s)]&=&\dot U_L^*(t,s)U_L(t,s)+U_L^*(t,s)\dot U_L(t,s)
\nonumber\\
\nonumber\\
&=&
iU_L^*(t,s)V_L(t,s)U_L(t,s)-iU_L^*(t,s)V_L(t)U_L(t,s)=0,~~~~~~~t,s\in\R.
\eeqn
Therefore, $U_L^*(t,s)U_L(t,s)=U_L^*(s,s)U_L(s,s)=I$. Finally, the operator $U_L(t,s)$  is invertible by (\re{Utsr})
with $r=t$.\bo
\bc
Any strong solution $C(\cdot)\in X$ to 
(\re{Hfr2}) admits the representation $C(t)=U_L(t)C(0)U_R(t)$
by the uniqueness of the  solution. Respectively, 
any strong solution $w(\cdot)\in X$ to 
(\re{RHFw}) admits the representation
\be\la{rep}
w(t)=U_0(t)U_L(t)U_0^*(t)w(0)U_0^*(t)U_R(t)U_0(t),~~~~~~~~~~~~~~~t\in\R.
\ee
\ec
Now the norm conservations (\re{nc}) obviously hold since 
all the operators $U_0(t)$, $U_L(t)$ and $U_R(t)$ are unitary.\bo

\section{A priori estimates and global solutions}

The conservation  laws imply the following a priori estimates.

\bl\la{lapres}
Let $w(\cdot)\in X_\ve$ be a strong solution to equation
(\re{RHFw}) for $|t|<\ve$ with an $\ve>0$. Then
\beqn\la{apres1}
\Vert w(t)\Vert_{\bH^1} &\le& C_1,~~~~~~~~~~~~~~~~~~~t\in(-\ve,\ve).
\\
\nonumber
\\
\la{apres2}
\Vert w(t)\Vert_{\bH^2} &\le& C_2 e^{C_3|t|},~~~~~~~~~~~~~t\in(-\ve,\ve).
\eeqn
where the constants $C_1$, $C_2$, and $C_3$ depend only on $\Vert w(0)\Vert_{\bH^2}$.

\el
\Pr We follow the scheme of \ci[Section 3.4]{CB1999}:
\medskip\\
i)  The first estimate follows from the energy conservation (\re{etrwr2}) since the last term
is nonnegative while the operator $H_0$ generates the Sobolev norm $H^1$.
\medskip\\
ii) The second estimate follows from the integral equation (\re{HFr3}). Namely, 
\be\la{HFr32}
 \Vert w(t)\Vert_{\bH^2}\le  C  [\Vert w(0)\Vert_{\bH^2}+\int_0^t \Vert\{\ti V(s),w(s)\}\Vert_{\bH^2}ds],~~~~~~~~~t\in(-\ve,\ve).
\ee
Now using (\re{Fw6}), we obtain
\be\la{HFr33}
 \Vert w(t)\Vert_{\bH^2}\le 
 C  [\Vert w(0)\Vert_{\bH^2}+C_F\int_0^t\Vert w(s)\Vert_{\bH^1}^2
 \Vert w(s)\Vert_{\bH^2}]ds,~~~~~~~~~t\in(-\ve,\ve).
\ee
Hence, (\re{apres2}) follows by the Gronwall lemma and  (\re{apres1}).\bo
\medskip\\
{\bf Proof of Theorem \re{t1}}. Lemmas \re{lls} and \re{lapres} imply Theorem \re{t1} by standard 
arguments.\bo

\setcounter{equation}{0}

\section{Agreement with the density matrix  formalism}

Let us discuss the  agreement of  the 
wave-matrix picture with the density matrix  formalism. 
First of all, the basic quantities (\re{dmm2}) coincide when $w(t)$ is  Slater-type 
density matrix
$w(t)=\sum_1^N|\psi_k(t)\rangle\langle\psi_k(t)|$ with the constraints (\re{cons}). In this case also $\ti H(t)=H(t)$.

Moreover, the density matrix  (\re{dmaw}) is invariant with respect to the transformation 
$w\mapsto Uw$ with any unitary operator $U$ in $X$.

Further let us consider separately the static and dynamic aspects.
\medskip\\
{\bf Static aspects.} 
Next lemma means the complete agreement
between the 
wave-matrix and the density-matrix formalism  in the ground state problem. 

\bl 
 The ground state energy (\re{gsHFL}) in the density-matrix theory and the wave-matrix picture
 coincide:
\be\la{gs2}
E^{HF}=\min\{\ti\cE^{HF}(w):  \Vert w\Vert\le 1,~\tr ww^*=N \}.
\ee

\el 
\Pr
i) (\re{gs2}) follows from (\re{gsHFL})
since 
$\ti\cE^{HF}(w)=\cE^{HF}(\ti K)$ 
by (\re{etrw}) and (\re{etr}), 
where 
$\ti K:=\fr12[ww^*+w^*w]\ge 0$, and $\tr \ti K:=\tr ww^*=\tr w^*w=N$.
\bo
\medskip\\
{\bf Dynamical aspects.} 
\bl\la{lK}
Let $w(\cdot)\in X$ be a strong solution to (\re{RHFw}) with 
$\tr w(0)w^*(0)=N$
and $\Vert w(0)\Vert\le 1$. Then 
the properties 
(\re{dmtr}) hold for the density matrices  $\ti K(t):=\fr12[w(t)w^*(t)+w^*(t)w(t)]$ and $K(t):=w(t)w^*(t)$ for all $t\in\R$.
\el
\Pr
Obviously, 
\be\la{dmtr2}
\ti K(t)\ge 0,~~~~ K(t)\ge 0.
\ee
Further, $\tr\ti  K(t)=\tr  K(t)$, and we know from
Lemma \re{lcc} and its proof that 
\be\la{dmtr3}
\tr \ti K(t)=\tr  K(t)=N
\ee
if $\tr K(0)=N$. It remains to note that
\be\la{KK}
\Vert \ti K(t)\Vert\le 1,~~~~\Vert K(t)\Vert\le 1,~~~~~~~~t\in\R
\ee
since
 $\Vert w(t)\Vert\le 1$ 
by (\re{nc}). 
\bo
\medskip

Let us recall in conclusion  that
the dynamics (\re{HFw}) for the wave matrix $w(t)$ agrees with the von Neumann
equation (\re{etr2}) for the corresponding density matrix $K(t):=w(t)w^*(t)$, see Remark \re{rvN}.

\setcounter{equation}{0}
\appendix

\section{Energy variation in wave-matrix picture} 
We prove Lemma \re{lGD}.
The G$\rm \hat a$teaux differentiability 
of the energy $\ti\cE^{RHF}(w)$ for $w\in \bH^2$ follows by the 
same arguments which justify the differentiation in time (\re{ecd}).
Hence, to justify (\re{GD}), it suffices to
differentiate formally each term on the right hand side of (\re{etrwr}). 
Additionally, we will differentiate also  the "exchange term" of (\re{etrw}).
\medskip\\
I.  For the first term $I_1=-\fr 14 [\langle\De_x w(x,y),w(x,y) \rangle +\langle\De_y w(x,y),w(x,y) \rangle $ the variation is obvious: 
\be\la{var1}
D_{\ov w(x,y)} I_1=-\fr12 [\De_x w(x,y)+\De_y w(x,y)],
\ee
which is the integral kernel of the anticommutator $\fr12\{-\De, w\}$.
\medskip\\
II.  For the second term $I_2=\ds\fr e4   \int \int
[V_n(x)+V_n(y)]| w(x,y)|^2dxdy$ the variation is also obvious: 
\be\la{var1n}
D_{\ov w(x,y)} I_2=\fr e2[V_n(x)+V_n(y)] w(x,y),
\ee
which is the integral kernel of the anticommutator $\fr12\{eV_n, w\}$.
\medskip\\
III.  For the third term $I_3=\ds\fr14 \ds\int\int \fr{ \ti\rho(x')\ti\rho(y')}{|x'-y'|}dx'dy'$ the  variation reads
\be\la{var2}
D_{\ov w(x,y)}I_2=\fr12 \int\int \fr{\ti\rho(y')}{|x'-y'|}D_{\ov w(x,y)} \ti\rho(x')dx'dy'.
\ee
Definition (\re{dmm}) implies that
\beqn\la{var3}
D_{\ov w(x,y)} [\ti\rho(x')]&=&\fr e2 D_{\ov w(x,y)} \int [w(x',z)\ov{w(x',z)}+ \ov{w(z,x')}w(z,x')]dz
\nonumber\\
\nonumber\\
&=&e [\de(x'-x)w(x',y)+ \de(x'-y)w(x,x')]
\eeqn
Substitution into (\re{var2}) gives
\beqn\la{var4}
D_{\ov w(x,y)} I_2&=&\fr e2\int\int \fr{\ti\rho(y')}{|x'-y'|}
[\de(x'-x)w(x',y)+ \de(x'-y)w(x,x')]dx'dy'
\nonumber\\
\nonumber\\
&=&\fr e2\int \fr{\ti\rho(y')}{|x-y'|}  w(x,y)dy'+ 
\fr e2\int\fr{\ti\rho(y')}{|y-y'|}w(x,y)dx',
\eeqn
which is the integral kernel of the anticommutator $\fr e2\{\ti V, w\}$, where the potential
$\ti V(x)$ is defined according to (\re{tiVt}):
\be\la{tiV}
\ti V(x)=\ds\int \fr{\ti\rho(y)}{|x-y|}dy.
\ee
IV. Similarly, 
for the exchange term, $I_4=-\ds\fr14 \ds\int\int\fr{|\ti\tau(x',y')|^2}{|x'-y'|}dx'dy'$,  the  variation reads
\be\la{var5}
D_{\ov w(x,y)} I_4=-\fr14\int\int \fr{
\ti\tau(x',y')D_{\ov w(x,y)} \ti\tau(y',x')
+\ti\tau(y',x')D_{\ov w(x,y)} \ti\tau(x',y')
}{|x'-y'|}dx'dy'
\ee
by (\re{dmm}).
Definition (\re{dmm}) implies that
\beqn\la{var6}
D_{\ov w(x,y)} \ti\tau(x',y')&=&\fr e2 D_{\ov w(x,y)} \int [ w(x',z)\ov{w(y',z)}+\ov{w(z,x')}w(z,y')]dz
\nonumber\\
\nonumber\\
&=&e [\de(y'-x)w(x',y)+ \de(x'-y)w(x,y')].
\eeqn
Substitution into (\re{var5}) gives
\beqn\la{var7}
D_{\ov w(x,y)} I_4=&-&\fr e4\int\int \fr{
\ti\tau(x',y')[\de(x'-x)w(y',y)+ \de(y'-y)w(x,x')]
}{|x'-y'|}dx'dy'
\nonumber\\
\nonumber\\
&-&\fr e4\int\int \fr{
\ti\tau(y',x') [\de(y'-x)w(x',y)+ \de(x'-y)w(x,y')]   
}{|x'-y'|}dx'dy'
\nonumber\\
\nonumber\\
=&-&\fr e4\int\int \fr{
\ti\tau(x,y')w(y',y)  
}{|x-y'|}dy'
-\fr e4\int\int \fr{
\ti\tau(x',y) w(x,x')
}{|x'-y|}dx'
\nonumber\\
\nonumber\\
&-&\fr e4\int\int \fr{
\ti\tau(x,x') w(x',y)
}{|x'-x|}dx'
-\fr e2\int\int \fr{
\ti\tau(y',y) w(x,y')   
}{|y-y'|}dy'
\nonumber\\
\nonumber\\
=&-&\fr e2\int\int \fr{
\ti\tau(x,y')w(y',y)  
}{|x-y'|}dy'
-\fr e2\int\int \fr{
\ti\tau(x',y) w(x,x')
}{|x'-y|}dx',
\eeqn
which is the integral kernel of the anticommutator $\fr e2\{\ti\cT, w\}$ where the operator 
$\ti\cT$ is defined according to (\re{tiVt}).\bo

%%%%%%%%%%%%%%%%%%%%%%%%%%%%%%%%%%%%%%%%%%%%%%%%%%%%%%%%%%%%%%%%%%%%%%%%%%%%%%%%%%%%%%

\end{document}